\documentclass[usenatbib,usegraphicx]{mn2e}

\usepackage{amsmath}
\usepackage{subfigure}
\usepackage{rotating}

\title[Multiple planet systems with SuperWASP]{A SuperWASP search for additional transiting planets in 24 known systems}

\author[A. M. S. Smith et al.]{A. M. S. Smith$^{1}$\thanks{E-mail:
amss@st-and.ac.uk},
L. Hebb$^{1}$,
A. Collier Cameron$^{1}$,
D. R. Anderson$^{2}$,
\newauthor
T. A. Lister$^{3}$,
C. Hellier$^{2}$,
D. Pollacco$^{4}$,
D. Queloz$^{5}$,
I. Skillen$^{6}$
and R. G. West$^{7}$\\
$^{1}$SUPA (Scottish Universities Physics Alliance), School of Physics \& Astronomy, University of St. Andrews, North Haugh,\\ 
St. Andrews, Fife, KY16 9SS\\
$^{2}$Astrophysics Group, Keele University, Staffordshire, ST5 5BG\\
$^{3}$Las Cumbres Observatory, 6740 Cortona Dr. Suite 102, Santa Barbara, CA 93117, USA\\
$^{4}$Astrophysics Research Centre, School of Mathematics \& Physics, Queen's University, University Road, Belfast BT7 1NN\\
$^{5}$Observatoire de Gen\`{e}ve, Universit\'{e} de Gen\`{e}ve, 51 Chemin des Maillettes, 1290 Sauverny, Switzerland\\
$^{6}$Isaac Newton Group of Telescopes, Apartado de Correos 321, E-38700 Santa Cruz de la Palma, Tenerife, Spain\\
$^{7}$Department of Physics and Astronomy, University of Leicester, Leicester LE1 7RH\\
}

\begin{document}
\maketitle

\newcommand{\hos}{{\sc hunt1star}}
\newcommand{\Mj}{\hbox{$M_{J}$\,}}
\newcommand{\Rj}{\hbox{$R_{\mathrm{J}}$\,}}
\newcommand{\Ms}{\hbox{$M_{*}$\,}}
\newcommand{\Rs}{\hbox{$R_{*}$\,}}
\newcommand{\rhos}{\hbox{$\rho_{*}$\,}}
\newcommand{\ns}{\hbox{$n_{*}$\,}}
\newcommand{\Mp}{\hbox{$M_{p}$\,}}
\newcommand{\Rp}{\hbox{$R_{\mathrm{p}}$\,}}
\newcommand{\Msol}{\hbox{M$_{\odot}$\,}}
\newcommand{\Rsol}{\hbox{R$_{\odot}$\,}}
\newcommand{\rhosol}{\hbox{$\rho_{\odot}$\,}}
\newcommand{\Me}{\hbox{M$_{E}$\,}}
\newcommand{\rmsub}[2]{#1_{\rm #2}}
\newcommand{\tick}{\ding{52}}
\newcommand{\cross}{\ding{56}}

\begin{abstract}

We present results from a search for additional transiting planets in 24 systems already known to contain a transiting planet. We model the transits due to the known planet in each system and subtract these models from lightcurves obtained with the SuperWASP survey instruments. These residual lightcurves are then searched for evidence of additional periodic transit events. Although we do not find any evidence for additional planets in any of the planetary systems studied, we are able to characterise our ability to find such planets by means of Monte Carlo simulations. Artificially generated transit signals corresponding to planets with a range of sizes and orbital periods were injected into the SuperWASP photometry and the resulting lightcurves searched for planets. As a result, the detection efficiency as a function of both the radius and orbital period of any second planet, is calculated. We determine that there is a good ($> 50$ per cent) chance of detecting additional, Saturn-sized planets in $P \sim$ 10 d orbits around planet-hosting stars that have several seasons of SuperWASP photometry. Additionally, we confirm previous evidence of the rotational stellar variability of WASP-10, and refine the period of rotation. We find that the period of the rotation is $11.91 \pm 0.05$ d, and the false alarm probability for this period is extremely low $(\sim 10^{-13})$.
\end{abstract}

\section{Introduction}
\subsection{Multiple planet systems}

Of the 347 presently known extra-solar planets, 90 are known to reside within multiple planet systems\footnote[8]{http://www.exoplanet.eu, 11th May 2009}. All of these systems, however, have been discovered by radial velocity measurements alone; none of them were discovered via the transit method, nor have any been later discovered to transit their host stars. The study of multiple planet systems enables greater understanding of theories of planet formation and migration, and affords us the opportunity to study planetary dynamics in action, as well as helping us answer fundamental questions such as `how common is the Solar System?'.

Planets that transit their host stars allow us to measure fundamental properties such as the planetary radius and remove much of the uncertainty on the value of the planetary mass by constraining the orbital inclination angle. This is just as true for multiple-planet systems, and further properties such as dynamical evolution time-scales can be measured for transiting systems \citep{Fab}.

\subsection{Detecting transiting multiple planet systems}

It has been noted that, given that there are now in excess of 50 known transiting planets, there is a good chance that at least one of these systems may harbour additional planets which we should be able to detect \citep{Fab}. There are three methods for detecting further planets in known transiting systems \citep{Fab}, namely (i) searching for transit timing variations (TTV) or variations in other transit parameters; (ii) searching for long-term radial velocity trends; and (iii) searching for additional transits.

In a multi-planet system, one planet can have a perturbing effect upon the orbit of a second planet, the effects of which can include small variations in the timings of transits caused by the second planet, such that the transits are no longer spaced periodically (\citealt{Agol05}; \citealt{Holman&Murray}). Many searches for further planets have been conducted, and continue to be conducted, using TTV, which has great sensitivity to planets in resonant orbits with the first planet, even if the second planet has an extremely low mass. 

It is also possible to infer the presence of additional planets by measuring long-term trends in other transit parameters. For instance, \cite{Coughlin} report observed increases in the orbital inclination, transit width and transit depth of Gl~436~b, which may indicate the presence of another planet.

In general, the discovery lightcurves of transiting planets, such as those produced by SuperWASP, are of insufficient quality to measure transit timings and other parameters with the required precision to discover additional planets; predicted timing variations, for instance, are typically on the order of seconds or tens-of-seconds. Additional resources must therefore be expended to obtain high precision lightcurves.

Secondly, known planets continue to be monitored for long-term trends in the radial velocity data, as is common practice with planets discovered by that means alone. Many longer-period additional planets have been found around stars around which a relatively short-period planet has been discovered by radial velocity measurements. Long-term monitoring of RV systems has yielded planets with periods of several years (the longest such period is 14.3 yr). This too, requires a modest expenditure of telescope time in order to obtain radial velocity data points at suitable intervals to detect longer-period planets.

Finally, photometric monitoring of known transiting systems may reveal transits caused by a second planet in the system. The inclination angle of this second planet must be sufficiently close to that of the first planet, that the second planet can be seen to transit its star as well as the first. This method, unlike the other two, does not necessarily require the allocation of further observing resources; instead the data archives of transit surveys can be searched for such transit signals. Such surveys often observe the same fields for several seasons, and so have a large quantity of data on known transiting systems. In this paper we present results from a search of the data archive of SuperWASP (Wide Angle Search for Planets), one such wide-field transit survey \citep{pollaccopasp}.

\subsection{Detecting transiting multiple planet systems by transit photometry}

Only two of the 58 known (to 2009 March) transiting planets have orbital periods, $P$, greater than 10 d (these were both detected initially by radial velocity means and were only later discovered to transit); indeed only six transiting planets have periods longer than 5 d. This is largely due to the selection effects present in wide-field surveys: (i) in general, a relatively large number of transits are required to boost the signal-to-noise sufficiently that the transit may be detected in the presence of correlated (`red') noise \citep{Smith1}. This requirement for many transits leads to the preferential detection of short-period planets that exhibit frequent transits. (ii) The probability that a given planetary system exhibits transits is inversely proportional to the orbital semi-major axis, $a$, and so this again causes a bias towards the detection of short period planets.

Any additional transiting planet is likely to orbit with a period greater than the currently known planet. Several factors militate against the usual difficulties in detecting long-period ($P >5$ d) planets, however. Most significantly, the probability that any second planet will transit should be greatly enhanced by the fact that there is already one transiting planet in the system. This is because the orbits of exoplanets in multiple systems are generally predicted to be close to coplanar, because such systems are believed to form from a flat disc in a similar fashion to the Solar System (e.g. \citealt{plan_book}). In the Solar System, this results in all planetary orbits being coplanar to within a few degrees.

If we assume co-planarity to within $5\degr$, the probability that the second planet transits is approximately equal to the ratio of the semi-major axes of the inner and outer planets, $a_\mathrm{in} / a_\mathrm{out}$ \citep{Fab}. This means, for example, the probability of a planet orbiting a solar analogue with a 10 day period is increased from about 5 per cent to around 45 per cent if an inner transiting planet exists with a period of 3 days.

Some previous attempts have been made to detect additional transiting planets. Croll et al. (2007a,b) used the {\it MOST} (Microvariability and Oscillations of STars) satellite to place upper limits on the presence of additional transiting planets in two systems (HD 209458 and HD 189733) known to harbour a transiting exoplanet. They were able to rule out the presence of additional planets larger than about 0.2 \Rj orbiting with periods less than 14 d for HD 209458 and planets larger than about 0.15 \Rj orbiting with periods less than 7 d for HD 189733.

This paper reports results from an extensive search of archival SuperWASP data for additional transits of stars known to host transiting exoplanets. We provide constraints on the existence of additional transiting planets in such systems.

\section{Observations}
\label{sec:obs}

Time-series photometry of 24 stars which host transiting exoplanets was obtained by the SuperWASP instruments, which are wide-field, multi-camera survey instruments described in \cite{pollaccopasp}.

Fifteen of these planets were observed by SuperWASP-N, located at the Roque de los Muchachos Observatory on La Palma in the Canary Islands, and nine by WASP-South, at the South African Astronomical Observatory in Sutherland. The objects observed are the first eighteen planetary systems discovered by SuperWASP (with the exception of WASP-9, for which follow-up observations are still ongoing) and seven similar systems discovered by other transiting planet surveys and retrospectively detected in SuperWASP data. Details of all these objects are given in Table \ref{tab:limits}. SuperWASP has an extensive archive of data on these objects. Many of them have been monitored for several observing seasons, dating back to 2004 in some cases.

\begin{table*}
\centering
\caption{Planetary systems searched for additional transiting bodies. The instrument used to observe the system is indicated in column 2: 'N' represents SuperWASP-N on La Palma and 'S' represents WASP-South in South Africa. Columns 3 to 10 are the parameters of the known planet used to subtract the transits of the first planet. They are the orbital period, $P$, the epoch of mid-transit, $\rmsub{t}{0}$, the transit duration, $w$, the impact parameter, $b$, the depth of the transit, $(\Rp / \Rs)^2$, the stellar density, \rhos, the orbital inclination angle, $i$, and the planetary radius, \Rp. Also shown is the number of data points in the lightcurve, $\rmsub{n}{phot}$.}

\label{tab:limits}
\begin{tabular}{lcccccccccl}
\hline
Star  & Obs.  & \multicolumn{8}{c}{MCMC fitted parameters used to subtract transits from lightcurves}     & \\
  & & $P$ /d & $\rmsub{t}{0}$ /HJD - & $w$ /d    & $b$  & $(\Rp / \Rs)^2$ & \rhos / \rhosol & i/$\degr$  & $ \rmsub{n}{phot}$ & Discovery reference\\
 &&&2450000&&&&&&&\\ 
\hline
WASP-1  &N& 2.519951 & 3998.19239 & 0.1551 & 0.056 &  0.01004 & 0.38709 & 89.44 & 13630 & \cite{wasp1_2}\\
WASP-2  &N& 2.152227 & 3978.60091 & 0.0736 & 0.729 &  0.01742 & 1.51264 & 84.80 &  7941 & \cite{wasp1_2}\\
WASP-3  &N& 1.846833 & 4214.03159 & 0.1112 & 0.446 &  0.01067 & 0.59680 & 85.20 &  5167 & \cite{wasp3}\\
WASP-4  &S& 1.338232 & 4387.32776 & 0.0881 & 0.055 &  0.02366 & 1.29600 & 89.44 & 10112 & \cite{wasp4}\\
WASP-5  &S& 1.628428 & 4373.99598 & 0.0987 & 0.314 &  0.01180 & 0.88028 & 86.85 & 15003 & \cite{wasp5}\\
WASP-6  &S& 3.361010 & 4593.07139 & 0.1068 & 0.204 &  0.02071 & 1.70136 & 88.96 & 14083 & \cite{wasp6}\\
WASP-7  &S& 4.954746 & 4133.65826 & 0.1555 & 0.183 &  0.00538 & 0.67368 & 89.02 & 24547 & \cite{wasp7}\\
WASP-8  &S& 8.158754 & 4679.33741 & 0.1457 & 0.332 &  0.01088 & 1.33364 & 88.67 & 16773 & Queloz et al. (in prep.)\\
WASP-10 &N& 3.092718 & 4299.09114 & 0.0944 & 0.225 &  0.02501 & 2.32879 & 88.94 &  8546 & \cite{wasp10}\\
WASP-11 &N& 3.722464 & 4473.05587 & 0.1066 & 0.065 &  0.01620 & 1.89252 & 89.70 &  8367 & \cite{wasp11}\\
&&&&&&&&&& \& \cite{hat10}\\
WASP-12 &N& 1.091423 & 4506.79316 & 0.1168 & 0.395 &  0.01390 & 0.33880 & 82.24 &  7233 & \cite{wasp12}\\
WASP-13 &N& 4.352999 & 4491.61656 & 0.1647 & 0.653 &  0.00886 & 0.28496 & 84.92 &  8661 & \cite{wasp13}\\
WASP-14 &N& 2.243738 & 4555.56986 & 0.1145 & 0.488 &  0.00979 & 0.61661 & 85.06 &  6817 & \cite{wasp14}\\
WASP-15 &S& 3.752087 & 4577.19424 & 0.1527 & 0.456 &  0.00953 & 0.45322 & 86.65 & 20442 & \cite{wasp15}\\
WASP-16 &S& 3.118600 & 4578.19167 & 0.0796 & 0.817 &  0.01208 & 1.13688 & 85.00 & 12237 & Lister et al. (ApJ, submitted)\\
WASP-17 &S& 3.735456 & 4566.65221 & 0.1777 & 0.103 &  0.01534 & 0.40353 & 89.21 & 16084 & Anderson et al. (in prep.)\\
WASP-18 &S& 0.941453 & 4600.88702 & 0.0895 & 0.336 &  0.00871 & 0.64012 & 84.47 &  8593 & Hellier et al. (in prep.)\\
HAT-P-4 &N& 3.056544 & 4099.10073 & 0.1707 & 0.040 &  0.00616 & 0.33246 & 89.63 &  6546 & \cite{hat4}\\
HAT-P-5 &N& 2.787765 & 3871.02389 & 0.1220 & 0.019 &  0.00626 & 0.83380 & 89.86 &  9081 & \cite{hat5}\\
HAT-P-6 &N& 3.853041 & 3931.62934 & 0.1648 & 0.054 &  0.00568 & 0.46098 & 89.61 & 20718 & \cite{hat6}\\
HAT-P-7 &N& 2.204799 & 4010.72625 & 0.1544 & 0.029 &  0.00258 & 0.30042 & 89.65 &  3647 & \cite{hat7}\\
TrES-2  &N& 2.470625 & 4011.98837 & 0.0750 & 0.864 &  0.01615 & 0.95940 & 83.46 &  6268 & \cite{tres2}\\
TrES-4  &N& 3.553964 & 4063.88015 & 0.1218 & 0.434 &  0.00769 & 0.84448 & 87.31 & 13790 & \cite{tres4}\\
XO-4    &N& 4.123971 & 4477.65872 & 0.1488 & 0.050 &  0.00680 & 0.68468 & 89.70 &  2589 & \cite{xo4}\\

\hline
\end{tabular}

\medskip
\end{table*}

\section{Search for additional planets}
\label{sec:hunt}

To search the SuperWASP lightcurves for additional transits we first take all existing lightcurves on a particular object from the SuperWASP archive. Several lightcurves may exist if the star has been observed in several seasons and/or with more than one camera. 

A systematic re-analysis of each of these systems was performed using all the photometric data either publicly available or held by the WASP consortium. The transits caused by the known planet were modelled using the Markov-chain Monte Carlo (MCMC) technique described by \cite{Cameron-etal07}, which uses the analytic model of \cite{M&A} (see Table \ref{tab:limits} for details of the parameters fitted). The resulting fitted parameters are within the joint error bars of the best published parameters. The modelled transits were then subtracted from the lightcurves, and the resulting residual lightcurves concatenated into a single residual lightcurve for each object.

We then search each of these residual lightcurves using \hos, a modified version of the adapted Box Least-Squares (BLS) algorithm used for routine SuperWASP transit hunting (\citealt{Cameron-etal06}; \citealt{Kovacs-etal02}). The \hos ~routine searches the lightcurve of a single object for transits, and it is able to handle large quantities of photometric data spanning multiple seasons and cameras, with a finely-sampled period grid.

\begin{figure}
\includegraphics[angle=270,width=8.25cm]{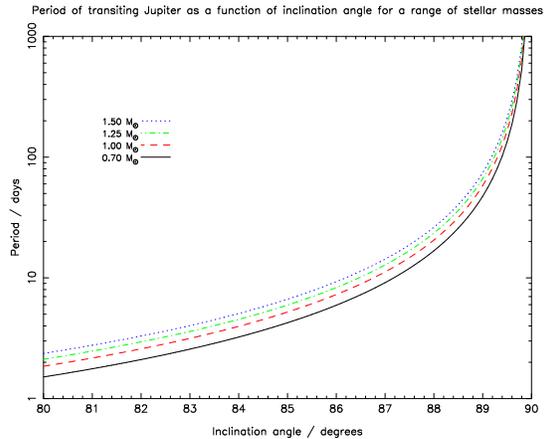}
\caption{Plot indicating the maximum period for a planet to exhibit transits, as a function of orbital inclination angle for a range of stellar masses. The host star is assumed to be on the main sequence and the planet is assumed to be Jupiter-sized.
}
\protect\label{fig:Pvi}
\end{figure}

The lightcurves of a total of 24 transiting planet host stars (see Table \ref{tab:limits} for the details of these objects) were searched for additional transits with periods of between 5 and 25 days using \hos. This lower limit of 5 d was chosen because, with the exception of WASP-8 b with a period of 8.16 d, all of the original planets have periods between 0.9 and 5.0 d, and so have been searched at sub-5\,d periods previously. The upper limit of 25 d was chosen after consideration of the length of the observing baselines and because for this period, the orbital inclination angle, $i$, must be greater than about 88\degr in order for transits to be exhibited by a Jupiter-sized planet (Fig. \ref{fig:Pvi}). The probability of such a planet transiting is around 2 -- 3 per cent assuming nothing about the system, and about 25 per cent if co-planarity to within 5 degrees with a $P = 3$ d planet is assumed.

A periodogram is produced for each object, and the parameters of the five strongest peaks in the periodogram are refined. The resulting candidates are loosely filtered according to the criteria usually used for SuperWASP planet hunting \citep{Cameron-etal06}. These criteria are (i) At least two transits must be detected; (ii) the reduced $\chi^2$ of the model must be less than 2.5; (iii) the phase-folded lightcurve must not consist of a high proportion of gaps; and (iv) the signal-to-red-noise ratio (the $S_\mathrm{red}$ statistic defined by \cite{Cameron-etal06}) must be -5 or less. We do not apply more strict filtering since we have a small number of objects, so the risk of false positives is small.

\section{Results of search for additional planets}
\label{sec:results}

Of the five best peaks for each of the 24 planets (120 peaks in total), 75 pass the criteria outlined in Sec. \ref{sec:hunt}, but our initial rejection criteria lean strongly towards retaining candidates.

\begin{figure}
\includegraphics[angle=270,width=8.25cm]{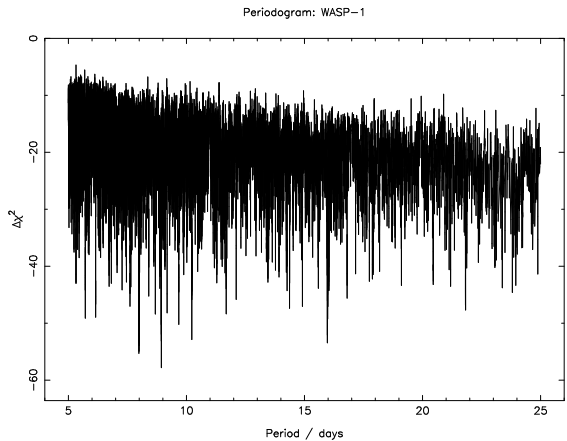}
\caption{Periodogram output of \hos ~for the residual lightcurve of WASP-1, after the subtraction of transits due to WASP-1~b. This is typical of the periodograms which exhibit no strong peaks.}
\protect\label{fig:w1per}
\end{figure}

\begin{figure}
\includegraphics[angle=270,width=8.25cm]{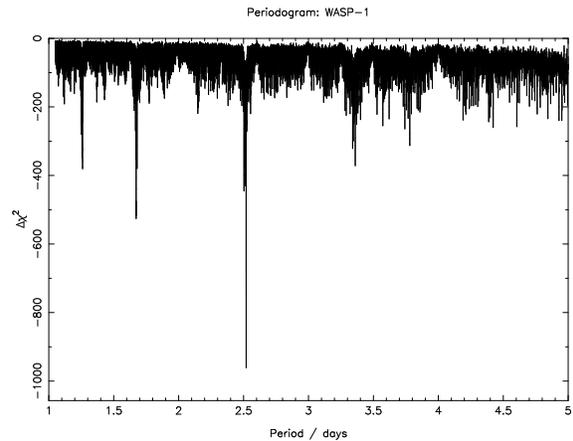}
\caption{Periodogram output of \hos ~for the unadulterated SuperWASP lightcurve of WASP-1. This is typical of the periodogram indicating the presence of a transiting planet, in this case WASP-1~b, which orbits with a period of about 2.52 d.}
\protect\label{fig:realper}
\end{figure}

As we are dealing with a small number of objects, we choose not to apply further cuts on, for example signal-to-noise, but instead visually inspect all 24 periodograms and all 120 phase-folded lightcurves. The periodograms produced for each of the 24 stars were inspected for any strong peaks indicative of a genuine transit, and  the phase-folded lightcurves were checked for transit-like signals. Most of the objects do not display any noteworthy periodogram peaks; Fig. \ref{fig:w1per} is a typical such periodogram, whereas Fig. \ref{fig:realper} is a periodogram typical of a genuine transiting planet. None of the phase-folded lightcurves display any transit-like signal.

\begin{figure}
\includegraphics[angle=270,width=8.25cm]{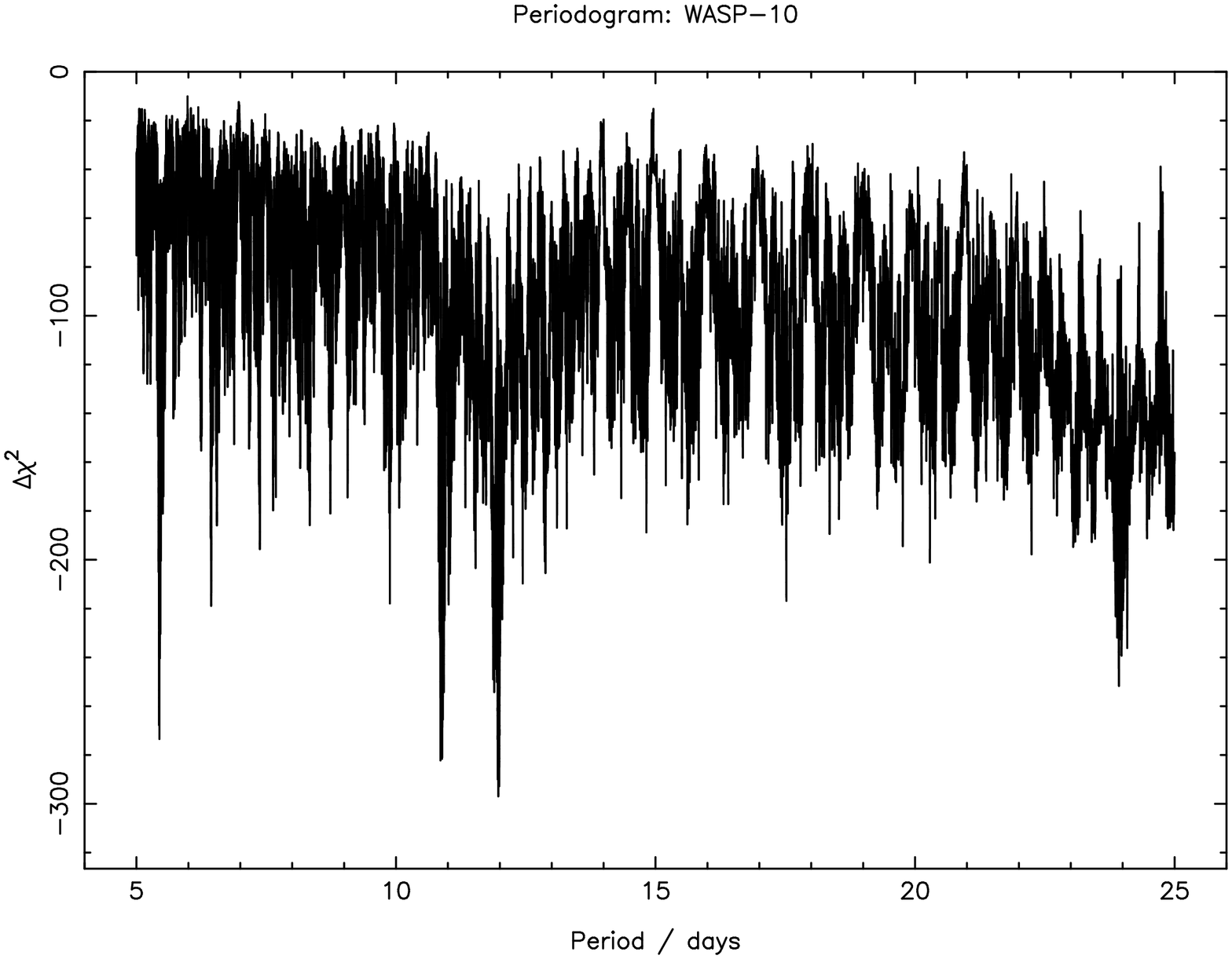}
\caption{Periodogram output of \hos~for WASP-10. The strong peak observed at about 12 d is caused by stellar rotation with that period.}
\protect\label{fig:w10per}
\end{figure}

Five phase-folded lightcurves were also plotted for each object, corresponding to each of the five strongest peaks in the periodogram. These were inspected for signs of a transit at phase 0, but no such signals were observed, even in the handful of objects that exhibit at least one reasonably strong periodogram peak.

\subsection{WASP-10}
\protect\label{sec:wasp10rot}

\begin{figure}
\includegraphics[angle=270,width=8.25cm]{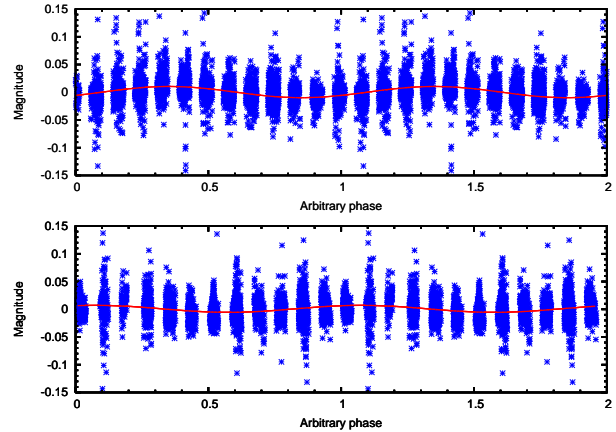}
\caption{Stellar rotation of WASP-10. The 2004 data are phased on a period of 11.949984 d (upper panel) and the 2006 data on a period of 11.946772 d (lower panel). Overplotted in each case is the best-fitting model of the form $\delta + A \mathrm{sin}(\omega t + \theta)$; the values of $\delta$, $A$ and $\theta$ are given in Table \ref{tab:w10_rot}.}
\protect\label{fig:w10_sine}
\end{figure}

Although a strongish peak at about 12 d is observed in the periodogram for WASP-10 (Fig. \ref{fig:w10per}), no transit is seen in the corresponding phase-folded lightcurve. This peak is less well-defined than the typical peak produced by planetary transits (Fig. \ref{fig:realper}). Instead, we attribute the peak to stellar rotation, the period, $P_{\mathrm{rot}}$, of which is known to be about 12 d \citep{wasp10}. We confirm the stellar rotation hypothesis by fitting a sine curve of the form $\delta + A \mathrm{sin}(\omega t + \theta)$, where $\omega = 2 \pi / P_{\mathrm{rot}}$ to the data (Fig. \ref{fig:w10_sine}). The best-fitting parameters for data taken in SuperWASP-N's 2004 and 2006 observing seasons are given in Table \ref{tab:w10_rot}. We find the same period of rotation (11.95 d) in both seasons of data, but both the phase of rotation and the amplitude of the variability differ between the two seasons.  We also compute the auto-correlation function \citep{E&K} of the data; the period determined by this method is 11.84 d.

Using a generalised Lomb-Scargle periodogram, as described in \cite{Z&K}, we are able to calculate the false alarm probability for the signal detected by sine fitting. Despite our lightcurve consisting of 8546 points, because of red noise the effective number of independent data points, $n_{\mathrm{eff}}$, is significantly smaller. To calculate $n_{\mathrm{eff}}$, the individual nights of data are shuffled, destroying the rotation signal, but maintaining the red noise. By assuming the highest peak in the resulting periodogram has a probability of 0.5, we calculate $n_{\mathrm{eff}} = 949$. The number of independent frequencies is calculated to be 1135, according to the approximation of \cite{Cumming04}. Using these values, and equation 24 of \cite{Z&K}, we calculate the false alarm probability to be $3.8 \times 10^{-13}$.

This extremely low probability confirms the reality of this variation, as does the fact that very similar periods were detected in different seasons of data, and with different methods. Taking the mean of all the periods detected, we conclude that the star rotates with $P_{\mathrm{rot}} = 11.91 \pm 0.05$. This variability is likely to be caused by starspots; such variability is not unusual amongst K-dwarfs.

\begin{table}
\centering
\caption{Best fitting parameters of a sine curve fitted to the lightcurve of WASP-10. The parameters are described in the text.}
\label{tab:w10_rot}
\begin{tabular}{lcccc}
\hline
&$\delta$/mmag & $A$/mmag & $P_{\mathrm{rot}}$/d & $\theta$/rad \\
\hline
2004 data & 0.3 & 10.1 & 11.949984 & -0.6351 \\
2006 data & 0.8 & 6.3 & 11.946772 & 1.0950 \\
\hline
\end{tabular}
\medskip
\end{table}

\section{Monte Carlo simulations}
\label{sec:sims}

In conclusion, none of the 24 lightcurves we analysed show any evidence for further transiting planets. Although no additional planets were detected in Sec. \ref{sec:results}, it is a useful exercise to quantify our ability to detect such planets, and hence to determine upper limits to the sizes and orbits of planets we have ruled out. To do this, we test our ability to detect planetary transits with various parameters, through the use of Monte Carlo simulations.

\subsection{Generation of artificial lightcurves}

The two parameters that most affect the ability of a survey such as SuperWASP to detect a planet are the size of the planet and the period of its orbit. We therefore choose to determine our ability to detect additional planets as a function of these two parameters, following an approach similar to that used by the {\it MOST} team to place upper limits on the presence of additional companions in the HD 209458 and HD 189733 systems (\citealt{Croll_209}a; \citealt{Croll_189}b).

We take the residual lightcurves described in section \ref{sec:hunt} and inject artificial transits into them, using the small planet approximation of \cite{M&A}. The inputs to the \cite{M&A} model are the stellar mass, stellar radius, effective stellar temperature ($\rmsub{T}{eff}$), stellar limb-darkening parameters and the planet radius and orbital inclination angle. We use values of \Ms, \Rs, and $\rmsub{T}{eff}$ taken from the exoplanet.eu website and the non-linear limb-darkening coefficients of \cite{Claret}, and we adopt an orbital inclination of 89\degr. Model planetary transits are injected at a range of 15 different orbital periods and 10 different planetary radii, giving a grid of 150 models (see Table \ref{tab:grid} for all values of $P$ and \Rp). One hundred different lightcurves are created for each of these points in $P-\Rp$ space, each with a randomly generated epoch of mid-transit, $t_\mathrm{0}$. 

\begin{table*}
\centering
\caption{Model planet parameters used in simulations. Ten different planetary radii and fifteen periods give a total of 150 combinations.
}
\label{tab:grid}
\begin{tabular}{ccccccccccccccc}
\hline
\multicolumn{15}{l}{Periods used / d:} \\
5.100 & 6.529 & 7.957 & 9.386 & 10.814 & 12.243 & 13.671 & 15.100 & 16.529 & 17.957 & 19.386 & 20.814 & 22.243 & 23.671 & 25.100 \\
\\
\multicolumn{15}{l}{Radii used / \Rj:} \\
0.40 & 0.55 & 0.70 & 0.85 & 1.00 & 1.15 & 1.30 & 1.45 & 1.60 & 1.75 & & & & & \\
\hline
\end{tabular}
\medskip
\end{table*}

\subsection{Searching for injected transits}

Each of the 15,000 lightcurves generated for each object is searched for transits by \hos, in exactly the same way as were the real data (Sec. \ref{sec:hunt}). The detection efficiency for a planet of a given size and orbital period is determined by the fraction of transits recovered for planets of those characteristics.

Whilst we inspected visually the periodograms and five best lightcurves of each of the 24 planets when searching the real data (Sec. \ref{sec:results}), this is clearly an unfeasible proposition for 15,000 lightcurves per object. Instead, for lightcurves identified as candidates by \hos, we require that the injected period, or an alias thereof, {\it and} the injected epoch of mid-transit are successfully recovered in at least one of the five best periodogram peaks. To do this, we define  the statistic,
\begin{equation}
\eta = \frac{|\rmsub{P}{meas} - \rmsub{P}{inj}|}{\rmsub{P}{inj}},
\end{equation}
where $\rmsub{P}{meas}$ and $\rmsub{P}{inj}$ are the orbital periods measured by \hos, and injected into the lightcurves, respectively. We then require $\eta < 0.001$ (corresponding to a detection at the injected period), or $0.4995 < \eta < 0.5005$ or $-0.0005 < \eta^* < 0.0005$, where $\eta^* = nint(\eta) - \eta$ (corresponding to detections at an alias of the injected period). These thresholds were designed to encapsulate clearly-defined populations of objects clustered around $\eta = 0$, $\eta = 0.5$ and $\eta = 1, 2,...$. Furthermore, we require that $\Delta t_0 \leq 0.10$ d, where $\Delta t_0 = \vert t_{0 inj} - t_{0 meas} \vert$, and $t_{0 inj}$ and $t_{0 meas}$ are the injected and measured epochs of mid-transit, respectively.

\begin{figure}
\centering
\subfigure{\label{fig:w4_sim_pgram}
\includegraphics[angle=270,width=8.25cm]{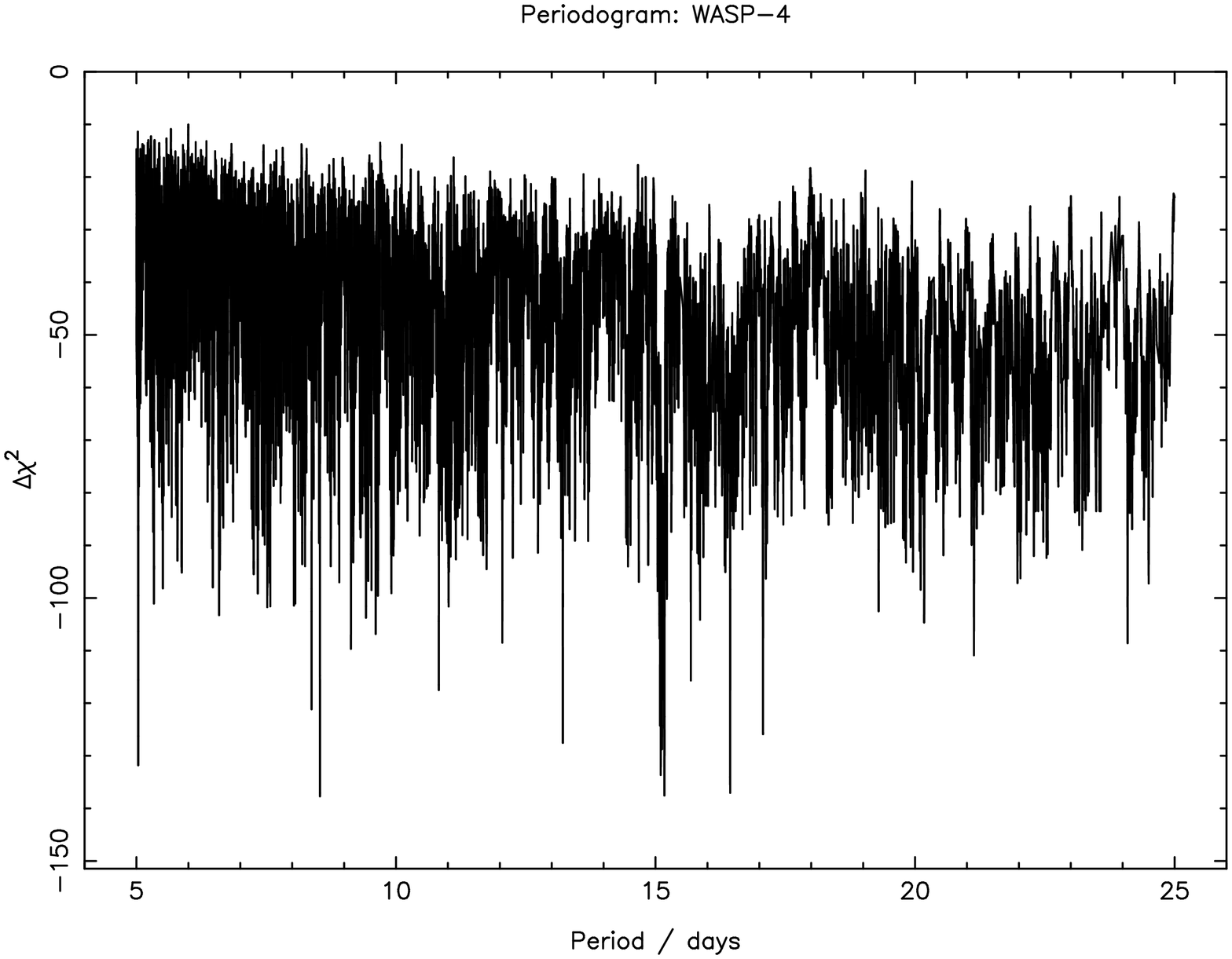}
}
\subfigure{\label{fig:w4_sim_lc}
\includegraphics[angle=270,width=8.25cm]{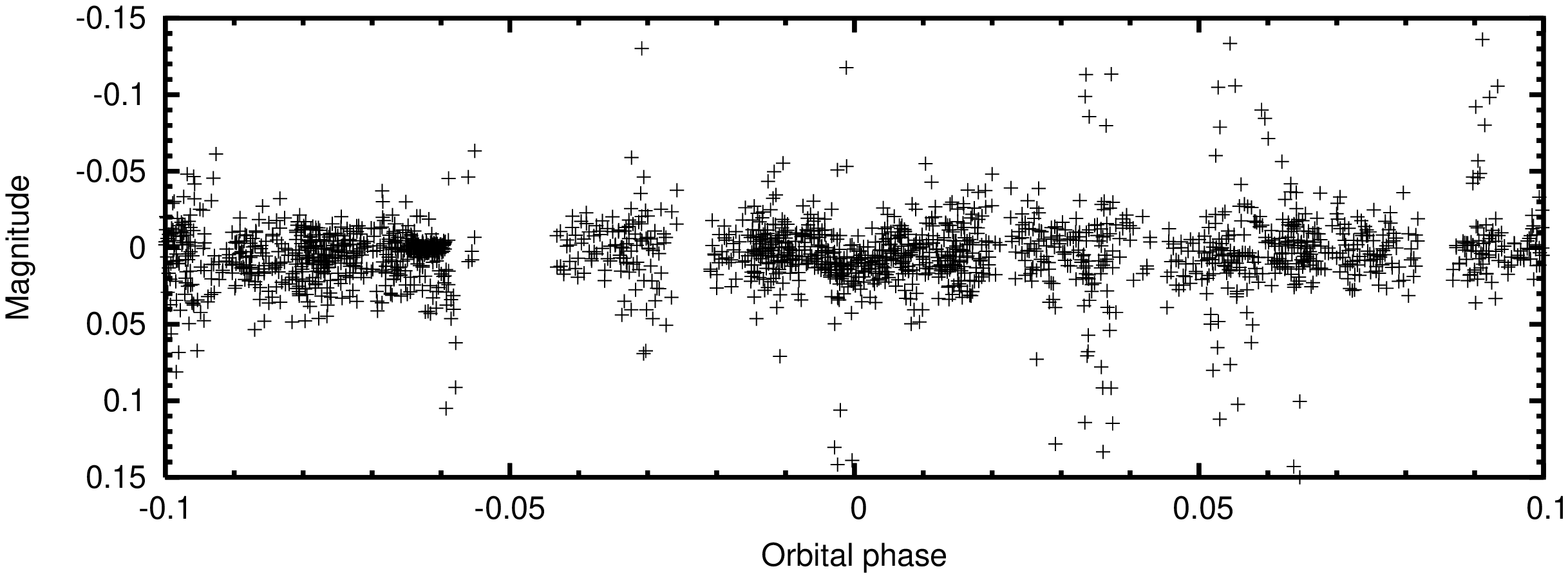}
}
\caption{An example of an artificial transit which is successfully recovered in our simulations. Shown are the \hos~results from transits corresponding to a planet with $P = 15.100$ d and $\Rp = 0.85 \Rj$, which were injected into the WASP-4 lightcurve. The planet is detected with $P = 15.10138$ d in the strongest peak of the periodogram (upper panel). The lightcurve, folded on the recovered period using the recovered epoch of mid-transit, exhibits a clear transit at phase = 0 (lower panel).}
\protect\label{fig:w4_sim}
\end{figure}

Periodograms and phase-folded lightcurves for several of the injected transits detected in this manner were inspected in the same fashion as the real data (Sec. \ref{sec:hunt}), in order to ensure that these objects could have been detected without prior knowledge of the orbital period. The periodogram and recovered lightcurve of one such injected transit is shown in Fig. \ref{fig:w4_sim}.

\subsection{Results of simulations}
\label{sec:sim_res}
\subsubsection{WASP-1}

\begin{figure}
\centering
\includegraphics[angle=270,width=7.25cm]{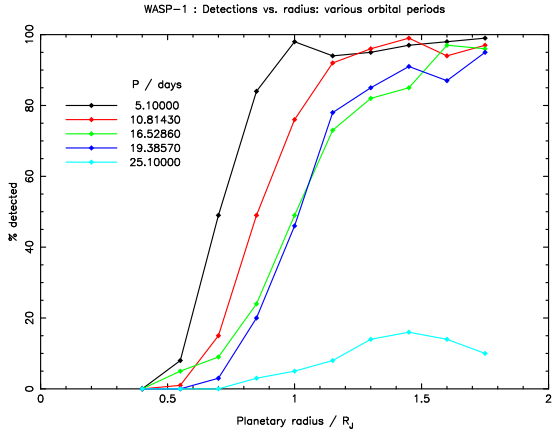}
\caption{WASP-1 simulation results (i). Detection efficiency as a function of planetary radius for a second planet orbiting WASP-1 for several orbital periods.}
\protect\label{fig:w1_det_r}
\end{figure}

\begin{figure}
\centering
\includegraphics[angle=270,width=7.25cm]{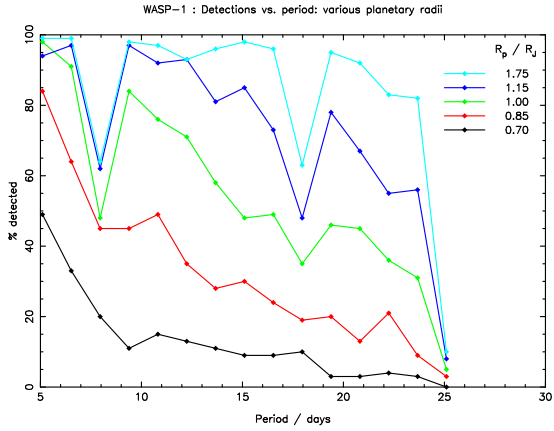}
\caption{WASP-1 simulation results (ii). Detection efficiency as a function of orbital period for a second planet orbiting WASP-1 for several planetary radii.}
\protect\label{fig:w1_det_p}
\end{figure}

The results of our simulations of extra planets in the WASP-1 system are presented in a series of 2 dimensional cuts through the parameter space (Figs. \ref{fig:w1_det_r} \& \ref{fig:w1_det_p}). Fig. \ref{fig:w1_det_r} shows detection efficiency plotted against the radius of the simulated additional planets for a variety of orbital periods, whereas in Fig. \ref{fig:w1_det_p}, detection efficiency is shown as a function of orbital period for a range of orbital periods.

\begin{figure}
\centering
\includegraphics[angle=0,width=7.25cm]{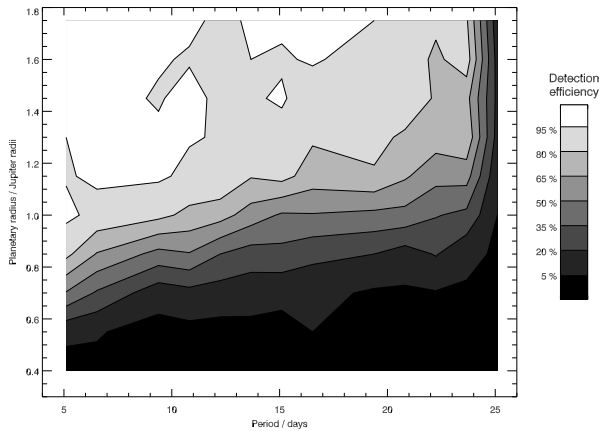}
\newline
\caption{WASP-1 simulation results (iii). Contour map showing detection efficiency as a function of orbital period and planetary radius.}
\protect\label{fig:w1_contour}
\end{figure}

The full dataset is shown as a contour map in Fig. \ref{fig:w1_contour}, although the results from the models with periods of 7.957 and 17.957 d are excluded as they are very close to an integer number of days. The sharp drops in detection efficiency observed at these periods (Fig. \ref{fig:w1_det_p}) are manifestations of the well-known 1 d alias phenomenon, which can cause either a sharp reduction or sharp increase in detection ability at periods which are almost exactly an integer number of days (e.g. \citealt{Smith1}).

\subsubsection{Other systems}

\begin{figure}
\centering
\includegraphics[angle=0,width=7.25cm]{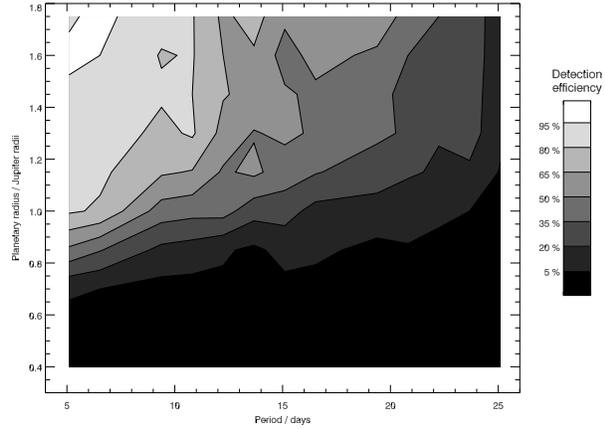}
\newline
\caption{HAT-P-4 simulation results. Contour map showing detection efficiency as a function of orbital period and planetary radius.}
\protect\label{fig:h4_contour}
\end{figure}

Similar simulations were conducted for several of the 24 systems which were searched for additional transits in Sec. \ref{sec:hunt}. Contour plots of the same type as Fig. \ref{fig:w1_contour} were also produced for these systems, although not all of them are shown here, for reasons of space. Our ability to detect additional planets in these other systems is very similar to that for WASP-1; there are variations in detection efficiency, but this correlates with the length of the original lightcurve. There is more data on WASP-1 than most of the other objects -- 13,630 data points spanning the 2004, 2006, and 2007 observing seasons. At the other extreme is HAT-P-4, which has a significantly shorter SuperWASP lightcurve, comprising just 6,546 data points, a few hundred of which are from 2006, with the rest from 2007. The contour plot for HAT-P-4 is shown in Fig. \ref{fig:h4_contour}, where it can be seen that the detection efficiency is poorer, particularly at relatively long periods, than for WASP-1 (Fig. \ref{fig:w1_contour}).

\section{Discussion}
\label{sec:discuss}

\subsection{Limits of simulations}

When considering the reliability of the upper limits established in Sec. \ref{sec:sim_res}, an obvious question to ask is `could systems in which multiple planets exhibit transits have been detected by SuperWASP (and HAT, TrES, and XO) in the first place?'. Although objects which exhibit photometric variability are generally rejected as planet candidates, we argue that systems with a sub-5\,d transiting planet and a further transiting planet with a period greater than 5 d would be unlikely to be rejected. In general, very few transits caused by the longer period planet would be present in the discovery lightcurve and that, combined with the shallow depth of planetary transits would prevent the candidate from failing test designed to eliminate variable stars. This is borne out by simulations conducted on lightcurves containing the transits caused by two planets, where the \hos~  algorithm was still able to detect the inner planet, despite the presence of additional transits.

In particular, we take the example of additional transits corresponding to a large, fairly short-period planet ($P = 12.243$ d; $\Rp = 1.45 \Rj$) which had been successfully recovered by \hos~ when injected into the residual WASP-4 lightcurve (Sec. \ref{sec:sims}). These same artificially generated transits were injected into the unmodified WASP-4~b lightcurve, in order to ascertain whether WASP-4~b could still be detected. WASP-4~b was indeed still detected, with only a very slightly reduced signal-to-red-noise value (-13.390 compared to -13.564).

\subsection{Lack of detections}

It has been suggested that the lack of transiting multiple planet systems is perhaps surprising given the prevalence of RV multiple systems. Specifically, \cite{Fab} notes that 11 companion planets are known to exist in the systems with the 33 shortest period planets. An apparently similar sample of transiting planets, however, contains no systems with known additional planets. Although \cite{Fab} does acknowledge the existence of observational biases which could partly explain this lack of transiting multiple planet systems, we argue that a careful examination of the characteristics of the known RV multiple planet systems reveals that the dearth of such systems is not surprising.

Of the 28 known multiple planet systems, only four have an inner planet which might reasonably have been detected by a wide-field transit survey, i.e. has $P <5$ d and $\Mp \mathrm{sin} i > 0.2~\Mj$, suggesting a radius large enough to be detected. None of these systems contains further planets with periods conducive to detection by means of transits; the second planet from the star orbits with $P > 95$ d in each of the four cases\footnote{The four systems are HIP 14810, Ups And, HD 187123, and HD 217107; the second planets in these systems orbit with periods of 95, 241, 3810, and 4210 d, respectively.}. In other words, there is currently no multiple-planet system of the kind we have demonstrated we are able to detect at present with SuperWASP known by RV studies. This does not, of course, mean that such systems do not exist, or that we should not look for them; especially given the low expenditure and potentially high reward involved.

\section{Conclusions}

We conducted a search for additional planets, with periods between 5 and 25 d, orbiting 24 stars known to harbour a transiting hot Jupiter, using SuperWASP photometry to search for additional transits. No planets were detected, so in order to place upper limits on the existence of such planets, we performed Monte Carlo simulations of planets of various sizes, and with various orbital periods. The results of these simulations suggest that, for objects like WASP-1 with three seasons of SuperWASP photometry, we have a good chance ($> 50$ per cent) of detecting Saturn-sized ($R_S = 0.843$ \Rj) planets out to about 10 d, and a sporting chance of detecting such planets with longer periods (there is a $\sim 20$ per cent chance of detecting a Saturn analogue in a 20 d orbit). These detection thresholds improve with increasing planet radius (up to about 1.2 \Rj), and are lower for stars, like HAT-P-4, with fewer data.

We are able to detect planets larger than about Saturn size, and with periods up to $\sim 20$ d with reasonable efficiency. As expected, the main factor affecting detectability is the time span of the lightcurve used to search for additional transits.

\subsection{Future prospects}

As our simulations (Sec. \ref{sec:sim_res}) demonstrated, the longer the span of the lightcurve, the greater our sensitivity to longer-period planets. As transit surveys continue to observe some of the known planets, our ability to detect additional transits in these systems will increase. Another possibility for increasing the number of available photometric data points is to share data between transit surveys, as suggested by \cite{Fleming08}.

\section{Acknowledgements}

AMSS acknowledges the financial support of a UK PPARC / STFC studentship. The SuperWASP Consortium consists of astronomers primarily from the Queen's University Belfast, St. Andrews, Keele, Leicester, The Open University, Isaac Newton Group La Palma, and Instituto de Astrof\'{i}sica de Canarias. The SuperWASP Cameras were constructed and operated with funds made available from Consortium Universities and the UK's Science and Technology Facilities Council. We acknowledge the use of software from the Condor Project (http://www.condorproject.org/) to run the Monte Carlo simulations. This research has made use of the NASA/IPAC/NExScI Star and Exoplanet Database, which is operated by the Jet Propulsion Laboratory, California Institute of Technology, under contract with the National Aeronautics and Space Administration.

\bibliographystyle{mn2e}
\bibliography{iau_journals,2planetsbib}
\bsp

\end{document}